\documentclass[aps,prc,twocolumn,showpacs,superscriptaddress,amsmath,amssymb,nofootinbib]{revtex4}
\usepackage{graphicx}% Include figure files
\usepackage{dcolumn}% Align table columns on decimal point
\usepackage{bm}% bold math
\begin{document}

%\preprint{APS/123-QED}

\title{Spectroscopy of $^{13}$B via the $^{13}$C($t$,$^{3}$He) reaction at 115 AMeV.}

\author{C. J. Guess}
\affiliation{National Superconducting Cyclotron Laboratory, Michigan State University, East
Lansing, MI 48824-1321, USA} \affiliation{Department of Physics and Astronomy, Michigan State
University, East Lansing, MI 48824, USA} \affiliation{Joint Institute for Nuclear Astrophysics,
Michigan State University, East Lansing, MI 48824, USA}
\author{R.G.T. Zegers}
\email{zegers@nscl.msu.edu} \affiliation{National Superconducting Cyclotron Laboratory, Michigan
State University, East Lansing, MI 48824-1321, USA} \affiliation{Department of Physics and
Astronomy, Michigan State University, East Lansing, MI 48824, USA} \affiliation{Joint Institute for
Nuclear Astrophysics, Michigan State University, East Lansing, MI 48824, USA}
\author{B.A. Brown}
\affiliation{National Superconducting Cyclotron Laboratory, Michigan State University, East
Lansing, MI 48824-1321, USA} \affiliation{Department of Physics and Astronomy, Michigan State
University, East Lansing, MI 48824, USA} \affiliation{Joint Institute for Nuclear Astrophysics,
Michigan State University, East Lansing, MI 48824, USA}
\author{Sam M. Austin}
\affiliation{National Superconducting Cyclotron Laboratory, Michigan State University, East Lansing, MI 48824-1321, USA}
\affiliation{Joint Institute for Nuclear Astrophysics, Michigan State University, East Lansing, MI 48824, USA}
\author{D. Bazin}
\affiliation{National Superconducting Cyclotron Laboratory, Michigan State University, East
Lansing, MI 48824-1321, USA}
\author{C. Caesar}
\altaffiliation[Present address: ]{GSI Darmstadt, Helmholtz-Zentrum f\"{u}r Schwerionenforschung, D-64291, Darmstadt, Germany}
\affiliation{National Superconducting Cyclotron Laboratory, Michigan State University, East
Lansing, MI 48824-1321, USA}
\affiliation{Johannes Gutenberg Universit\"{a}t, Mainz, Germany}
\author{J.M. Deaven}
\affiliation{National Superconducting Cyclotron Laboratory, Michigan State University, East
Lansing, MI 48824-1321, USA} \affiliation{Department of Physics and Astronomy, Michigan State
University, East Lansing, MI 48824, USA} \affiliation{Joint Institute for Nuclear Astrophysics,
Michigan State University, East Lansing, MI 48824, USA}
\author{G.F. Grinyer}
\affiliation{National Superconducting Cyclotron Laboratory, Michigan State University, East
Lansing, MI 48824-1321, USA}
\author{C. Herlitzius}
\altaffiliation[Present address: ]{Physics Department, Technische Universit\"{a}t M\"{u}nchen, D-85748 Garching, Germany}
\affiliation{National Superconducting Cyclotron Laboratory, Michigan State University, East
Lansing, MI 48824-1321, USA}
\affiliation{Johannes Gutenberg Universit\"{a}t, Mainz, Germany}
\author{G.W. Hitt}
\affiliation{National Superconducting Cyclotron Laboratory, Michigan
State University, East Lansing, MI 48824-1321, USA} \affiliation{Department of Physics and
Astronomy, Michigan State University, East Lansing, MI 48824, USA} \affiliation{Joint Institute for
Nuclear Astrophysics, Michigan State University, East Lansing, MI 48824, USA}
\author{S. Noji}
\affiliation{Department of Physics, The University of Tokyo, Bunkyo, Tokyo 113-0033, Japan}
\author{R. Meharchand}
\affiliation{National Superconducting Cyclotron Laboratory, Michigan State University, East
Lansing, MI 48824-1321, USA} \affiliation{Department of Physics and Astronomy, Michigan State
University, East Lansing, MI 48824, USA} \affiliation{Joint Institute for Nuclear Astrophysics,
Michigan State University, East Lansing, MI 48824, USA}
\author{G. Perdikakis}
\affiliation{National Superconducting Cyclotron Laboratory, Michigan
State University, East Lansing, MI 48824-1321, USA} \affiliation{Joint Institute for
Nuclear Astrophysics, Michigan State University, East Lansing, MI 48824, USA}
\author{H. Sakai}
\affiliation{Department of Physics, The University of Tokyo, Bunkyo, Tokyo 113-0033, Japan}
\author{Y. Shimbara}
\affiliation{Graduate School of Science and Technology, Niigata University, Niigata 950-2181, Japan}
\author{C. Tur}
\affiliation{National Superconducting Cyclotron Laboratory, Michigan
State University, East Lansing, MI 48824-1321, USA} \affiliation{Joint Institute for
Nuclear Astrophysics, Michigan State University, East Lansing, MI 48824, USA}

\date{\today}%

\begin{abstract}
Gamow-Teller and dipole transitions to final states in $^{13}$B were studied via the $^{13}$C($t$,$^{3}$He) reaction at $E_{t}=115$ AMeV. Besides the strong Gamow-Teller transition to the $^{13}$B ground state, a weaker Gamow-Teller transition to a state at 3.6 MeV was found. This state was assigned a spin-parity of $3/2^{-}$ by comparison with shell-model calculations using the WBP and WBT interactions which were modified to allow for mixing between $n\hbar\omega$ and $(n+2)\hbar\omega$ configurations. This assignment agrees with a recent result from a lifetime measurement of excited states in $^{13}$B. The shell-model calculations also explained the relatively large spectroscopic strength measured for a low-lying $1/2^{+}$ state at 4.83 MeV in $^{13}$B. The cross sections for dipole transitions up to $E_{x}(^{13}$B)$=20$ MeV excited via the $^{13}$C($t$,$^{3}$He) reaction were also compared with the shell-model calculations. The theoretical cross sections exceeded the data by a factor of about 1.8, which might indicate that the dipole excitations are ``quenched". Uncertainties in the reaction calculations complicate that interpretation.

\end{abstract}

\maketitle

\section{Introduction}
\label{sec:intro}
The study of neutron-rich nuclei near mass number A=12 has received considerable attention because of experimental evidence for the presence of states with non-zero $\hbar\omega$ configurations at low excitation energies. Good examples are the ground state of $^{11}$Be, which has spin-parity $J^{\pi}=1/2^{+}$ \cite{WIL59} and the low-lying (2.7 MeV) $1^{-}$ state in $^{12}$Be \cite{IWA00}. Such effects have been attributed to changes in the neutron shell structure. A variety of mechanisms has been proposed, such as the deformation of these neutron-rich nuclei \cite{SHI07,HAM07} or effects related to the monopole component of the tensor force \cite{SUZ03}.

Ota {\it et al.} \cite{OTA08} identified a low-lying $1/2^{+}$ state at 4.83 MeV in $^{13}$B in a $^{12}$Be($\alpha,t$) experiment performed in inverse kinematics at 50 AMeV. The relatively large spectroscopic strength of $0.20\pm0.02$ found for this state (with a systematic error of 60\%)  could not be reconciled with shell-model calculations in the $psd$ model space, which took into account excitations up to 3$\hbar\omega$ with an interaction designed to include the tensor effects \cite{SUZ03}. It was suggested that the low-lying $1/2^{+}$ state is highly deformed and shifted to low excitation energies. Recently, Iwasaki {\it et al.} \cite{IWA09} found evidence for a long-lived state at 3.53 MeV in $^{13}$B, with a suggested $J^{\pi}$ of $3/2^{-}$. The associated low transition strength indicates that the state has a dominant $\nu 2p 2h$ configuration and that the $N=8$ shell closure in $^{13}$B is weak.

In this work, excited states in $^{13}$B were investigated through the $^{13}$C($t$,$^{3}$He) charge-exchange reaction at 115 AMeV.
The ($t$,$^3$He) reaction selects isovector transitions ($\Delta T=1$) in the $\Delta T_{z}=+1$ direction (i.e. $(n,p)$-like) and, like other charge-exchange probes, provides an excellent tool to study the spin-isospin response of nuclei \cite{HAR01,OST92}. In the present analysis, we focused on Gamow-Teller (GT) ($\Delta L=0$, $\Delta S=1$) and dipole transitions ($\Delta L=1$, $\Delta S=0,1$). The goal was to gain further insight into the properties of states in $^{13}$B that are excited via the $^{13}$C($t$,$^{3}$He) reaction. The results were compared with shell-model calculations, which were also used to better understand the findings of Refs. \cite{OTA08,IWA09}

At 115 AMeV, the ($t$,$^{3}$He) reaction at forward scattering angles is predominantly of a single-step, direct nature. Under such circumstances, the cross section for GT transitions at zero linear momentum transfer ($q=0$) is proportional to GT strength (B(GT)) \cite{TAD87}:
\begin{equation}
\label{eq:eik}
\frac{d\sigma}{d\Omega}(q=0)=\hat{\sigma}B(GT),
\end{equation}
where $\hat{\sigma}$ is the unit cross section. The unit cross section can be calibrated by using a transition for which the B(GT) is known from the log$ft$ value obtained through $\beta$-decay experiments. For the transition from the ground state of $^{13}$C ($J^{\pi}=1/2^{-}$) to the ground state of $^{13}$B ($J^{\pi}=3/2^{-}$) a B(GT) of 0.711(2) can be derived from the measured $\beta$-decay half-life (log$ft$=4.034(6) \cite{AJZ91}). The proportionality of Eq. \ref{eq:eik} is not perfect, largely due to coherent $\Delta$L=2 contributions to the $\Delta$L=0 GT transitions, which are mediated via the non-central tensor interaction. Such effects have been studied experimentally and theoretically for a number of cases \cite{ZEG06,COL06,FUJ07,ZEG07,ZEG08}.

Whereas the proportionality between the differential cross section at $q=0$ and the transition strength has been well established for GT transitions, this is not the case for transitions involving larger units of angular momentum transfer, such as dipole transitions. In a theoretical study by Dmitriev {\it et al.} \cite{DMI01} for the $^{12}$C($p$,$n$) reaction, it was shown that an approximate proportionality with transition strength exists if the cross section at the peak region of the dipole differential cross section is used (i.e. at finite momentum transfer). Also based on theoretical studies, Yako {\textit{et al.}} \cite{YAK06} assumed a proportionality between the peak differential cross section (at finite angle) for the $^{90}$Zr($p$,$n$) and $^{90}$Zr($n$,$p$) reactions and dipole strength.
However, both studies rely on the similarity between transition densities for different excitations and a broader study involving several nuclei over a wide mass-range and experimental checks have not been performed. Therefore, in the present analysis of dipole excitations via the $^{13}$C($t$,$^{3}$He) reaction, no proportionality was assumed a-priori. Instead, experimental cross sections were compared directly with theoretical ones.

\section{Experiment and data extraction.}
\label{sec:experiment}
A primary beam of $^{16}$O was accelerated to 150 MeV/nucleon in the coupled K500 and K1200 cyclotrons \cite{CCF} at the National Superconducting Cyclotron Laboratory.  The beam impinged upon a 3526 mg/cm$^{2}$ Be production target and produced a 115 AMeV secondary triton beam \cite{HIT06}. A 195 mg/cm$^2$ Al wedge was placed at the intermediate image of the A1900 fragment separator \cite{MOR03} to rid the beam of its $^6$He contamination, which produced minor backgrounds in previous ($t,^3$He) experiments through the ($^6$He$\rightarrow$$^3$He+3n) breakup reaction.  In addition, a slit at the intermediate image was used to limit the momentum acceptance to $\frac{dp}{p}=\pm0.25\%$. The tritons were guided through the analysis beam line to a target placed at the object of the S800 spectrometer \cite{BAZ03}. To monitor the triton yield during the experiment for the purpose of absolute cross section measurements, the triton rates at the object of the S800 were measured by using a plastic scintillator and calibrated against the readout of a non-intercepting primary beam probe, located at the exit of the K1200 cyclotron. The continuous readout of the non-intercepting probe during regular data taking then provided a direct measure for the triton rate. The calibration procedure was carried out several times during the experiment (typically after beam tuning) to monitor small changes (less than 10\%) in the calibration factor.
The transmission of tritons from the focal plane of the A1900 to the object of the S800 was around 85\%, an improvement over previous experiments \cite{COL06,HOW08,HIT09} stemming from a realignment of the analysis beam-line between the A1900 and the S800. The improvement gave a yield at the target of 10$^7$ s$^{-1}$.

An 18 mg/cm$^{2}$, 99.3\% isotopically enriched $^{13}$CH$_2$ target \cite{NOJ08} was placed at the object of the S800 spectrometer.  The analysis beam-line to the S800 spectrometer was operated in dispersion-matched mode in order to improve the energy resolution. $^3$He particles were identified in the focal plane of the S800 \cite{YUR99} by using the $\Delta E-E$ signals from two plastic scintillators and the time difference between the cyclotron RF signal and the event trigger (an event in the first scintillator).  The dispersive and non-dispersive positions and angles of the $^3$He particles in the focal plane were determined by using two cathode readout drift detectors (CRDCs).  Dispersive and non-dispersive angles, the non-dispersive position, and the momentum of the $^3$He at the target were reconstructed by using a fifth-order \textsc{COSY Infinity} transfer matrix \cite{BER93} based on measured maps of the magnetic field.  The dispersion-matched beam spot size at the target was approximately 5 cm in the dispersive direction and 1 cm in the non-dispersive direction, but the large size of the target (7.5 cm by 2.5 cm) eliminated scattering from the frame.  A missing mass calculation was performed on an event-by-event basis to reconstruct the excitation energy of $^{13}$B.  The scattering angle was reconstructed from the non-dispersive and dispersive angles up to 5.75$^\circ$ in the center of mass; the angular resolution was 0.6$^\circ$ (FWHM).  The acceptance of the spectrometer depends not only on horizontal and vertical components of the scattering angle, but also on the hit-position in the non-dispersive direction at the target. Therefore, a Monte Carlo simulation was performed to estimate the acceptance of the spectrometer as a function of scattering angle and was used in the calculation of opening angles required for calculating differential cross sections. The resolution in the reconstructed excitation energy of $^{13}$B was 480 keV (FWHM) and was dominated by the difference in energy loss for tritons and $^{3}$He particles in the target.

\section{Experimental results}
\label{sec:results}
In Fig. \ref{fig1}(a) the measured $^{13}$C($t$,$^{3}$He) excitation energy spectrum is shown. Besides the dominant transition to the $^{13}$B ground state, several other distinct peaks were observed at excitation energies below 6 MeV. Above that energy, the spectrum is more complex. Several broader structures are observed that lie on top of a structureless continuum. The continuum is mostly due to quasifree processes with three-body final states (see e.g. \cite{MAT80,AAR81}).

To gain insight into the various contributions to the excitation energy spectrum, angular distribution were studied and compared with theory. For that purpose, differential cross sections were calculated in the distorted-wave Born approximation (DWBA), by using the code \textsc{FOLD} \cite{FOLD}. In this code, the effective nucleon-nucleon interaction by Love and Franey \cite{LOV81,LOV85} is double-folded over the transition densities of the projectile-ejectile ($t$-$^{3}$He) and target-residual ($^{13}$C-$^{13}$B) systems. A
short-range approximation as described in Ref. \cite{LOV81} was used for the exchange terms in the
potential. For $^{3}$He and $^{3}$H, densities used in the folding were obtained from Variational
Monte-Carlo results \cite{WIR05}. One-body transition densities (OBTDs) for the transitions from
$^{13}$C to $^{13}$B were calculated with the code OXBASH \cite{OXBA} employing the WBP interaction \cite{WAR92} in the $spsdpf$ model space. The Hamiltonians were modified as discussed in more detail in Section \ref{sec:theory}. Radial wave functions were calculated by using a Woods-Saxon potential. Binding energies of the particles were determined in OXBASH \cite{OXBA} by employing the Skyrme SK20 interaction \cite{BRO98}.
The optical potential parameters used in the DWBA calculation were extracted from $^{3}$He elastic scattering on $^{13}$C \cite{FUJ04}. The depths of the triton potentials were calculated by scaling the depths of the $^{3}$He potentials by 0.85, while leaving radii and diffusenesses constant \cite{WER89}.

\subsection{Gamow-Teller transitions}
\begin{figure}
\includegraphics[scale=0.9]{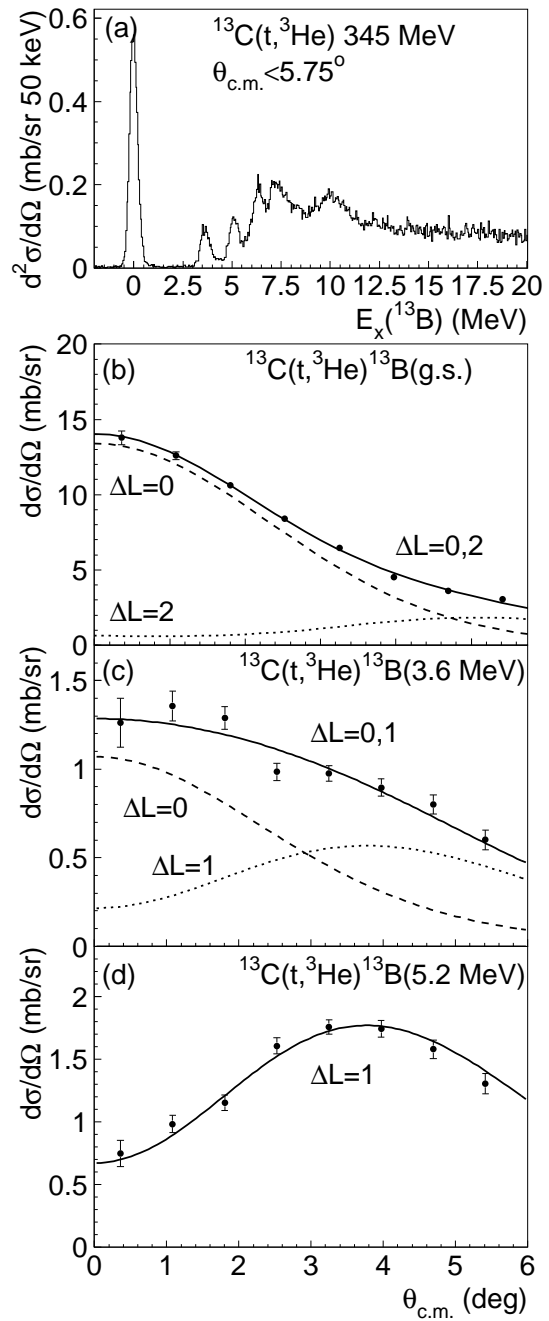}
\caption{\label{fig1} (a) Excitation-energy spectrum in $^{13}$B, measured via the $^{13}$C($t$,$^{3}$He) reaction at 115 AMeV. (b) Differential cross section for the transition to the $^{13}$B(g.s.) and the decomposition in $\Delta L=0$ (dashed line) and $\Delta L=2$ (dotted line) components. The solid line is the sum of the two components. (c) Differential cross section for the peak in the $^{13}$B spectrum at 3.6 MeV and the decomposition into $\Delta L=0$ (dashed) and $\Delta L=1$ (dotted) components. The solid line is the sum of the two components. (d) Differential cross section for the peak in the $^{13}$B spectrum at 5.2 MeV. The data is well described by a pure dipole transition. The theoretical cross section in (b), (c) and (d) were calculated in DWBA and scaled to match the data as discussed in the text.}
\end{figure}

The $^{13}$C($1/2^{-}$,g.s.)$\rightarrow$$^{13}$B($3/2^{-}$,g.s.) transition is associated with total angular momentum transfers $\Delta J=1$ and $\Delta J=2$, i.e. there are two distinct sets of OBTDs. The $\Delta J=1$ transition has two components: the Gamow-Teller component with $\Delta L=0$, $\Delta S=1$ and a quadrupole component with $\Delta L=2$, $\Delta S=1$.
The $\Delta J=2$ transition also has two components: one associated with spin transfer ($\Delta L=2$, $\Delta S=1$) and one without ($\Delta L=2$, $\Delta S=0$). To extract the cross section associated with the Gamow-Teller component of the transition from the data, the $\Delta L=0$, $\Delta S=1$ contribution has to be isolated, and this can be done by investigating the experimental angular distribution. Whereas the $\Delta L=0$, $\Delta S=1$ cross section peaks at $\theta_{cm}=0^{\circ}$, the angular distribution of the $\Delta L=2$ components are relatively flat at forward scattering angles, with a maximum at $\theta_{cm} \approx 5^{\circ}$.
In Fig. \ref{fig1}(b) the measured differential cross section of the $^{13}$C($1/2^{-}$,g.s.)$\rightarrow ^{13}$B($3/2^{-}$,g.s.) transition is shown and decomposed into the $\Delta L=0$ and $\Delta L=2$ components by using angular distributions generated in DWBA. Since the three contributions to the $\Delta L=2$ component have very similar angular distributions over the solid angle covered in the experiment, their relative strengths were kept fixed and their sum treated as a single contribution to the differential cross section. The normalizations of the $\Delta L=0$ and combined $\Delta L=2$ contributions were adjusted independently to best fit the experimental data. The extracted cross section for the $\Delta L=0$ component at $0^{\circ}$ was $13.1\pm1.3$ mb/sr, where the error is deduced from the uncertainties in the fit (0.2 mb/sr) and an estimated 10\% systematic uncertainty in the determination of absolute cross sections due to beam-intensity fluctuations and variations of the target-thickness over the large area covered by the impinging triton beam.

Since the application of Eq. \ref{eq:eik} requires the use of the differential cross section at $q=0$ (which corresponds to Q-value $Q=0$ and $\theta_{cm}=0^{\circ}$), the extracted cross section of the $\Delta L=0$ component at $0^{\circ}$, but finite $Q$, (for the $^{13}$C($t$,$^{3}$He)$^{13}$B(g.s) reaction, $Q=-13.41$) was extrapolated to $Q=0$ by multiplying with the ratio
$\left[\cfrac{d\sigma}{d\Omega}(Q=0,0^{\circ})/\cfrac{d\sigma}{d\Omega}(Q=-13.41,0^{\circ})\right]=1.24$. The differential cross sections in the numerator and the denominator were calculated in DWBA. Since the B(GT) for $^{13}$C($1/2^{-}$ g.s.)$\rightarrow ^{13}$B($3/2^{-}$ g.s.) is known, the unit cross section can be extracted from the cross section at $q=0$ and its value was $\hat{\sigma}_{exp}=22.8\pm2.3$ mb/sr. Within error bars, this value is consistent with the one found for the $^{13}$C($^{3}$He,$t$) reaction at 140 MeV/nucleon ($20\pm 1$) \cite{ZEG08}.
The unit cross section was also estimated in the theory, by dividing the DWBA cross section at $q=0$ by the theoretical B(GT). The value found was $27.3$ mb/sr, which is a factor of $1.20\pm 0.12$  higher than the experimental value. The discrepancy is likely due to the approximate treatment of exchange contributions in the theory \cite{UDA87,KIM00} and was also observed for the ($^{3}$He,$t$) reaction on variety of target nuclei \cite{ZEG07}. Systematic uncertainties related to the optical potential parameters are another possible source for the discrepancy.

Fig. \ref{fig1}(c) displays the differential cross section of the peak observed at $E_{x}$($^{13}$B)$\approx3.6$ MeV. Although the angular distribution peaks at forward angles, it is clear by comparison to the ground-state transition in Fig. \ref{fig1}(b) that this peak contains stronger component(s) not associated with $\Delta L=0$. Nakayama {\it et al.} reported \cite{NAK90} a strong ($\sim80$\% of the ground state transition) transition with $\Delta J^{\pi}=2^{-}$ to unresolved states at $E_{x}$($^{13}$B)$\approx3.5-4$ MeV via the $^{13}$C($^{7}$Li,$^{7}$Be) reaction at 21 MeV/nucleon. In Ref. \cite{GUI00}, evidence was found for $3/2^{+}$ and $5/2^{+}$ states at 3.48 and 3.68 MeV, respectively, by using one-neutron knockout from $^{14}$B. From these previous results, it is clear that the contribution of dipole transitions should be considered in the analysis of the peak at 3.6 MeV.
In earlier experiments aimed at investigating levels in $^{13}$B \cite{MID64,WYB72,AZJ78}, four states were found in the energy window between 3.4 and 3.8 MeV, two of which were associated with $\Delta L=1$, and two with $\Delta L=2$. In addition, similar to the ground state transition, the $\Delta J=1$ transfer associated with the GT transition also has a $\Delta L=2$ component.

Taking into account the possible contributions from the various $L$ transfers, the GT contribution to the peak at 3.6 MeV was extracted by fitting pairwise combinations of $\Delta L=0$, $\Delta L=1$ and $\Delta L=2$ angular distribution calculated in DWBA to the data. The peak observed at 5.2 MeV served as a test for how well the DWBA calculations could describe dipole transitions. As shown in Fig. \ref{fig1}(d), the shape of the theoretical angular distribution matches very well the experimental data (the theoretical curve was scaled). For the peak at 3.6 MeV, a fit with a combination of $\Delta L=0$ and $\Delta L=1$ angular distributions provided the best result, which is shown in Fig. \ref{fig1}(c). A fit with all three angular-momentum transfer components was performed as well, but the $\Delta L=2$ component was small and the reduced $\chi^{2}$ did not improve compared to the fit where the $\Delta L=2$ component was not included. The $\Delta L=0$ contribution to the differential cross sections changed by 5\% between the three-component fit and the fit with $\Delta L=0$ and $\Delta L=1$ components only. This percentage was used as an estimate for the systematic error in the extraction procedure. The extracted cross section at $0^{\circ}$ was $1.07\pm0.07\pm0.05$ mb/sr, where the first uncertainty is due to the statistical error and the second due to the systematic error. This cross section was extrapolated to $q=0$ and the B(GT) determined by using the unit cross section derived from the ground state transition. Its value was $0.065\pm0.005$, where the error includes the statistical and systematic uncertainties in the fit of the angular distribution. Following the procedure for estimating the systematic error in the extracted B(GT) due to interference between $\Delta L=0$ and and $\Delta L=2$ amplitudes presented for the analysis of the $^{13}$C($^{3}$He,$t$) reaction in Ref. \cite{ZEG08}, we found that such uncertainties were small ($<$2\% of the extracted B(GT) values) compared to the statistical and systematical (fitting) uncertainties and thus could be ignored.
No further unambiguous signatures of GT transitions to states at excitation energies above 4 MeV were found in the analysis of the $^{13}$B excitation energy spectrum. A summary of the results for the GT transitions to the ground state and the excited state at 3.6 MeV is provided in Table \ref{tab:gt}.

\begin{table}
\caption{\label{tab:gt}Summary of experimental results for Gamow-Teller transitions studied via the $^{13}$C($t$,$^{3}$He) reaction.}
\begin{ruledtabular}
\begin{tabular}{ccccc}
$E_{x}$($^{13}$B) & $[\frac{d\sigma}{d\Omega}(0^{\circ})]_{L=0}$ & $[\frac{d\sigma}{d\Omega}(q=0)]_{L=0}$ & B(GT) & $\hat{\sigma}$ \\
MeV & mb/sr & mb/sr &  & mb/sr \\
\hline \\
0 & $13.1(1.3)$ & $16.2(1.6)$ & $0.711(2)\footnotemark[1]$ & $22.8(2.3)$ \\
3.6 & $1.07(9)$ & $1.48(12)$ & $0.065(5)\footnotemark[2]$ & - \\
\end{tabular}
\end{ruledtabular}
\footnotetext[1]{Deduced from the measured log$ft$ value in $\beta$-decay \cite{AJZ91}.}
\footnotetext[2]{Deduced from the unit cross section determined from the ground-state transition.}
\end{table}

\subsection{Dipole transitions}
The extraction of the contributions from dipole transitions to the spectrum shown in Fig. \ref{fig1}(a) was complicated by the presence of the quasifree continuum and, possibly, isolated transitions associated with angular momentum transfer ($\Delta L\neq1$). The magnitude of the latter contributions were estimated by generating shell-model states up to 20 MeV excitation energy for transitions to final states with $J^{\pi}=1/2^{+},3/2^{+},5/2^{+},7/2^{+}$ and $J^{\pi}=1/2^{-},3/2^{-},5/2^{-},7/2^{-}$ and calculating their differential cross sections in DWBA. Above $E_{x}=4$ MeV, the contribution to the total cross section from transitions with $\Delta L\neq1$ amounted to maximally 5\% and the response was thus strongly dominated by the dipole transitions. This is due to the fact that the Gamow-Teller strength is mostly confined to low excitation energies as discussed above and transitions with $\Delta L>1$ are suppressed at these beam energies and forward scattering angles.
\begin{figure*}
\includegraphics[scale=1.0]{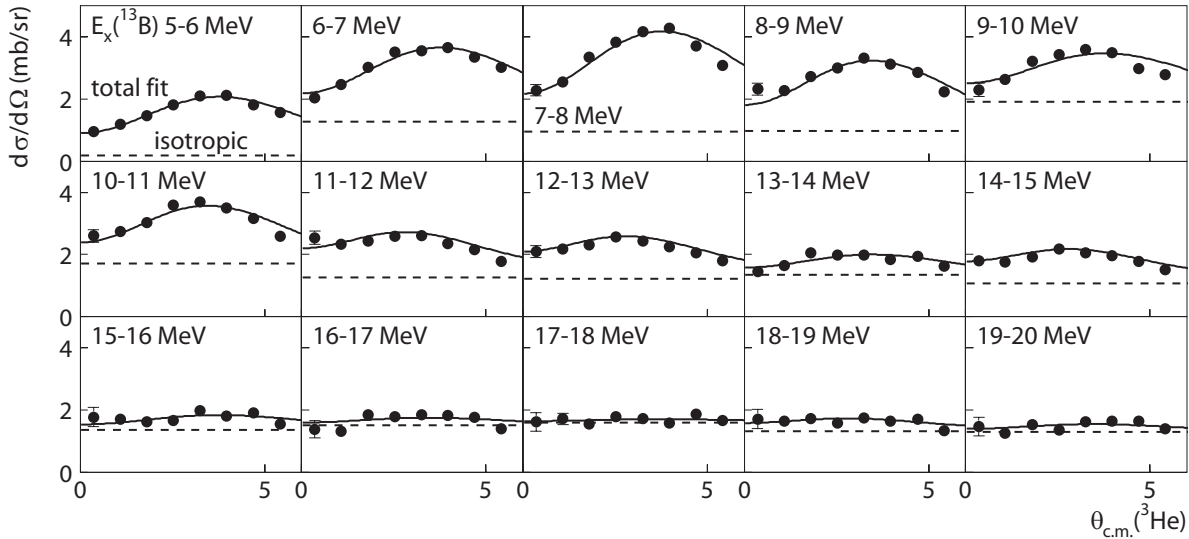}
\caption{\label{fig2} Decomposition of differential cross section measured for excitation energies in $^{13}$B between 5 and 20 MeV. The decomposition into a dipole and isotropic component is performed for 1-MeV bins in excitation energy. The isotropic component represents the continuum and small contributions from transitions with $\Delta L>1$. For details, see text.}
\end{figure*}
The quasifree continuum has a relatively flat angular distribution at forward scattering angles and, due to its nature, should not exhibit large local fluctuations as a function of excitation energy. In the case of the ($t$,$^{3}$He) reaction and ignoring minor contributions from multistep reactions (e.g. breakup-pickup), it is largely due to charge-exchange reactions on a bound proton and the ejection of the exchanged neutron, leaving the residual ``spectator" nucleus in its ground or an excited single proton-hole state \cite{JAN93}. The Fermi motion of the bound proton leads to a broadening of the energy of the outgoing $^{3}$He particle. The threshold for three-body breakup is the neutron separation energy ($S_{n}$($^{13}$B)=4.878 MeV) which thus defines the onset of the continuum. There are few detailed studies of the shape of the continuum and sometimes a phenomenological description has been used \cite{JAN93}, originally employed in $\pi$ charge-exchange reactions \cite{ERE86}. More detailed studies require coincidence measurements between the ejectile and the knocked-out particle (see e.g. \cite{MAT80,AAR81,BOR88,ZEG99}).
\begin{figure}
\includegraphics[scale=1.0]{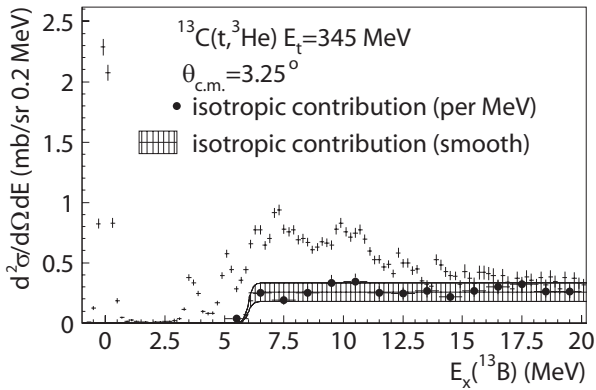}
\caption{\label{fig3}Excitation energy spectrum measured for $3.0^{\circ}<\theta_{c.m.}<3.5^{\circ}$. The isotropic contribution due to quasifree processes for each 1-MeV bin above 5 MeV is indicated by solid markers. The hatched band is the assumed contribution of the quasifree processes in the further analysis of the spectrum. The width of the band  indicates the estimated uncertainty in these contributions.}
\end{figure}
Since the contributions from non-dipole transitions were expected to be small, the excitation energy spectrum above 5 MeV was decomposed into two components: one stemming from collective dipole transitions, which are associated with angular distributions that exhibit a strong minimum at $0^{\circ}$ and peak at finite scattering angle, and one due to the continuum. The continuum was assumed to have a flat angular distribution, since it is a superposition of components associated with a wide range of angular momentum transfers (including $\Delta L=1$). Minor contributions from transitions to specific states associated with $\Delta L\geq1$ have rather flat angular distributions as well and they were effectively absorbed in the continuum component.
Since the continuum is smooth, the decomposition was performed for 1-MeV wide bins in excitation energy, as shown in Fig. \ref{fig2}. For each energy bin, the experimental differential cross sections are shown. The results of the fit with the flat and dipole contributions are superimposed (solid lines), as well as the contribution from the flat component only (dashed lines). Different sets of OBTDs for dipole transitions predicted in the shell-model were employed to calculate the angular distributions used in the fit. As the angular distributions did not vary much when switching from one set of OBTDs to another, the extracted dipole contributions to the experimental spectrum were not sensitive to the choice of the set of OBTDs used and the associated systematic errors were significantly smaller than those due to other effects (see below) and thus ignored.
Clear dipole contributions could be distinguished up to about 16 MeV. Beyond that energy, the angular distributions were nearly flat, indicating that the non-collective excitations associated with a range of angular momentum transfers dominated the response. In what follows, we use the term ``continuum" to denote this sum of non-collective excitations.

In Fig. \ref{fig3}, the excitation energy spectrum at center-of-mass scattering angles between 3$^\circ$ and 3.5$^\circ$ is shown. Superimposed are the isotropic contributions representing the continuum (per 1 MeV bin) extracted from the decomposition. Within the uncertainties, the continuum contribution is indeed rather smooth as a function of excitation energy. For the remainder of the analysis, the continuum was, therefore, assumed to be constant above 6 MeV, with the magnitude set to the average of the continuum contributions in each of the 1 MeV bins above that energy. Taking into account the assumptions made in the decomposition and the fact that small dipole contributions to the continuum might have been missed in the analysis, the uncertainty in this magnitude was chosen to equal the maximum spread in continuum contributions in each of the 1 MeV bins. Between 4.8 MeV (the onset of the continuum) and 6 MeV, a smooth function raising the continuum contribution from 0 to the average level was used. From Fig. \ref{fig3} it is clear that even though the uncertainties in the continuum contribution are large (about half the magnitude of the continuum itself), the extraction of cross sections associated with dipole transitions can be performed with relatively high accuracy at this scattering angle since they contribute so dominantly to the spectrum.

\section{Comparison with theory}
\label{sec:theory}
\subsection{Theoretical framework}
We calculate the energies and transition strengths with the WBP and WBT Hamiltonians \cite{WAR92}. These Hamiltonians
were derived from fits to experimental energy levels under the condition that the wave functions are described by pure $n\hbar\omega$ configurations. With this assumption, the lowest lying states of $^{13}$C and $^{13}$B
are obtained from the negative parity $0\hbar\omega$ configurations
$(0s)^{4}(0p)^{9}$. The 1$\hbar\omega$ excitations relate to excited positive parity states
of the form $(0s)^{3}(0p)^{10}$ and $(0s)^{4}(0p)^{8}(1s0d)^{1}$, where the
latter configuration dominates due to the relative small shell gap between $0p$ and $0s1d$ shells. The 2$\hbar\omega$ excitations would describe higher-lying negative parity states dominated by the $(0s)^{4}(0p)^{7}(1s0d)^{2}$ configurations. The energies of these states obtained with WBP and WBT are shown in Fig. 2 of Ref. \cite{GUI00}. Continuing this sequence we would have 3$\hbar\omega$ positive parity states dominated by $(0s)^{4}(0p)^{6}(1s0d)^{3}$
with the lowest state being a 1/2$^{ + }$ at 5.65 (5.73) MeV with WBP(WBT). This state is close in energy to a state at 4.83 MeV observed strongly in the $^{4}$He($^{12}$Be,$^{13}$B) transfer reaction \cite{OTA08}. In this picture, this 3$\hbar\omega$ 1/2$^{ + }$ state is the band-head of a collective band with other states at energies of (with WBP) 6.48 MeV for 5/2$^{ + }$ and 9.04 MeV for 9/2$^{ + }$. These states have a similar structure to
the $(1s0d)^{3}$ configurations (relative to a closed $p$-shell for $^{16}$O)
of $^{19}$F coupled to the $(0s)^{4}(0p)^{6}$ configurations of $^{10}$Be.

As discussed in Ref. \cite{NAV00}, the lowest 0$\hbar\omega$ and 2$\hbar\omega$ states
obtained with WBT in $^{12}$Be are degenerate in energy. The experimental data for neutron knockout \cite{NAV00} suggest that the ground state of $^{12}$Be has 32\% 0$\hbar\omega$ and 68\% 2$\hbar\omega$ (the results
labeled WBT2 in Ref. \cite{NAV00}). The spectroscopic factor calculated with
WBT for the pure 2$\hbar\omega$ (0$^{+}$) to 3$\hbar\omega$ 1/2$^{+}$ transfer is 0.33, giving a value of 0.22 when the WBT2 wave functions are used, in good agreement with the experimental value of 0.20(2) \cite{OTA08}. This good agreement is in contrast with the discrepancy found between theory and experiment as presented in Ref. \cite{OTA08} for reasons discussed next.

As emphasized in Ref. \cite{WAR92}, the WBP and WBT interactions are designed to be used for pure $n\hbar\omega$ configurations. But one can generate matrix elements which connect the $n\hbar\omega$ and $(n+2)\hbar\omega$
configurations from the potential parameters obtained with WBP. When this is used to mix the configurations, the relative spacing of the $n\hbar\omega$ parts of the spectrum can be greatly distorted. The reason, as discussed in Ref. \cite{WAR92a}, is due to the mixing of the many high-lying $(n+2)\hbar\omega$ configurations
with the low-lying $n\hbar\omega$ states via the $(\lambda\mu=20)$ SU3 tensor part of the $\Delta n=2\hbar\omega$
interaction that pushes down the energies of the $n\hbar\omega$ states.
Thus to be consistent one must also include $(n+4)\hbar\omega$ admixtures in order to push down the $(n+2)\hbar\omega$ states, etc. One would need to go to perhaps 8-10 $\hbar\omega$ for a consistent calculation. But at
the same time one should change the basic interaction since it is designed for the simple pure $n\hbar\omega$ configurations.
Thus, it is not correct to carry out mixed $(0+2)\hbar\omega$ and $(1+3)\hbar\omega$ calculations for $^{13}$B as was done in Ref. \cite{OTA08}.

An approximate way to correct for this distortion is discussed in \cite{WAR92a}. One calculates the energy of the
lowest state with $n\hbar\omega$ and with $n\hbar\omega +(n+2)\hbar\omega$ configurations.
The energy shift between these two results, $\Delta$E, can then be applied to lower all of the $(n+2)\hbar\omega$ configurations to implicitly take into account the effect that the $(n+4)\hbar\omega$ configurations have in lowering the energy of the $(n+2)\hbar\omega$ configurations. The energy shift obtained in this way for our $A=13$ case is about $\Delta E = 4$ MeV. With this approximation the lowest states in $^{13}$B with WBP dominated by each $n\hbar\omega$ are the 3/2$^{-}$ (76\% 0$\hbar\omega$) ground state, 1/2$^{ + }$ (78\%1$\hbar\omega$) at 2.90 MeV,
3/2$^{-}$ (94\% 2$\hbar\omega$) at 3.68 MeV and 1/2$^{ + }$ (61\%3$\hbar\omega$) at 5.73 MeV.
The results obtained for the wave functions and transition strengths are similar for the WBT Hamiltonian.
The 3/2$^{-}$ state predicted at 3.68 MeV likely corresponds to the state observed at 3.53 MeV in Ref. \cite{IWA09}.

With this mixed (0$+$2)$\hbar\omega$ wave function for $^{12}$Be the spectroscopic
factors for the $^{12}$Be to $^{13}$B transfer for the first five 1/2$^{ + }$
states at 2.90, 5.73, 7.11, 8.00 and 9.55 MeV are 0.034, 0.183, 0.030, 0.020 and 0.014, respectively.
Based on these results, we associate the second 1/2$^{ + }$ state with the ``proton intruder" level observed at 4.83 MeV with a strength of 0.20(2) in Ref. \cite{OTA08}.

The B(GT) values for transitions from $^{13}$C to low-lying negative parity states in $^{13}$B are dominated by transitions to the 3/2$^{-}$ (76\% 0$\hbar\omega$) ground state with B(GT) = 0.904 and to the 3.68 MeV 3/2$^{-}$ (94\% 2$\hbar\omega$)
state with B(GT) = 0.025. The B(GT) for the excited 3/2$^{-}$ state is directly related to the amount of $\hbar\omega$ mixing. For example, if the $^{13}$C ground state is constrained to be pure 0$\hbar\omega$, then the results are B(GT)$=0.990$ (0.073) to the first (second) 3/2$^{-}$ states in $^{13}$B. Thus, there
is destructive interference between the $\hbar\omega$ components of the transition to the second 3/2$^{-}$ state.
Matrix elements of the type $  <0\hbar \omega \mid GT\mid 2\hbar \omega >  $ are zero, and
the result for the first two 3/2$^{-}$ states can be understood schematically in terms of a three component model:
$$
\mid {\rm ^{13}C}>\, = 0.918 \mid {\rm ^{13}C} ,0\hbar \omega > + 0.397
\mid {\rm ^{13}C} ,2\hbar \omega >,
$$
$$
\mid {\rm ^{13}B}>_{1} = 0.871 \mid {\rm ^{13}B} ,0\hbar \omega > +
0.491 \mid {\rm ^{13}B} ,2\hbar \omega >_{1},
$$
$$
\mid {\rm ^{13}B}>_{2} = -0.237 \mid {\rm ^{13}B} ,0\hbar \omega > +
0.972 \mid {\rm ^{13}B} ,2\hbar \omega >_{2},
$$
$$
<{\rm ^{13}C} ,0\hbar \omega \mid GT\mid {\rm ^{13}B} ,0\hbar \omega
>\, = 1.147,
$$
$$
<{\rm ^{13}C} ,2\hbar \omega \mid GT\mid {\rm ^{13}B} ,2\hbar \omega
>_{1} = 0.674,
$$
and
$$
<{\rm ^{13}C} ,2\hbar \omega \mid GT\mid {\rm ^{13}B} ,2\hbar \omega
>_{2} = 1.059.
$$
In addition, the (2$\hbar\omega$) $\rightarrow$ (2$\hbar\omega$) matrix elements
are sensitive to the wave function (and interaction) as shown for the case of  $^{16}$O $  \rightarrow  $ $^{16}$F Gamow-Teller strength in \cite{SNO83}.

\begin{table*}
\caption{\label{tab:compare}Comparison of experimental results and theoretical predictions for Gamow-Teller transitions up to 5 MeV excited via the $^{13}$C($t$,$^{3}$He) reaction.}
\begin{ruledtabular}
\begin{tabular}{ccccccccc}
\multicolumn{2}{c}{experiment\footnotemark[1]} & \multicolumn{7}{c}{theory} \\
\cline{1-2} \cline{3-9} \\
\multicolumn{2}{c}{} & \multicolumn{1}{c}{} & \multicolumn{3}{c}{CKII-$0\hbar\omega$} & \multicolumn{3}{c}{WBP $(0+2)\hbar\omega$} \\
\cline{4-6} \cline{7-9} \\
$E_{x}(^{13}$B) & B(GT) & $J^{\pi}$ & $E_{x}(^{13}$B) (MeV) & B(GT) & B(GT) & $E_{x}(^{13}$B) & B(GT) & B(GT) \\
MeV & & & MeV & & quenched\footnotemark[2] & MeV & & quenched\footnotemark[2] \\
\cline{1-2} \cline{3-6} \cline{7-9} \\
0 & 0.711(2) & $3/2^{-}$ & 0 & 1.02 & 0.691 & 0 & 0.904 & 0.612 \\
3.6 & 0.065(5) & $3/2^{-}$ & - & - & - & 3.86 & 0.025 & 0.017 \\
 &  & $1/2^{-}$ & 4.59 & 0.00538 & 0.00386 & 3.675 & 0.00270 & 0.00183 \\
\end{tabular}
\end{ruledtabular}
\footnotetext[1]{See Table \ref{tab:gt}.}
\footnotetext[2]{A reduction factor of 0.677 is applied (for details see text).}
\end{table*}

\subsection{Comparison between theoretical and experimental results for Gamow-Teller transitions}

To compare the theoretical results for the Gamow-Teller strengths with the experimental values in Table \ref{tab:gt} one has to take into account that theoretical calculations for Gamow-Teller strengths in general overestimate the data due to degrees of freedom not included in the models. This reduction factor can be quantified by squaring the ``quenching" factor $q_{s}$ for the free-nucleon operator. For nuclei in the $p$ shell, $q_{s}=1-0.19(\frac{A}{16})^{0.35}$ \cite{CHO93}, which gives 0.823 for $A=13$ and a reduction of the Gamow-Teller strength by a factor $(0.823)^{2}=0.677$ is expected. It is incorrect to apply this reduction factor to the calculated strengths in the model in which mixed (0$+$2)$\hbar\omega$ wave functions are employed, since part of the quenching is thought to originate from such configuration mixing \cite{HYU80,ARI99,BRO88}. Therefore, applying the full reduction factor would constitute partial double-counting of quenching effects and the true Gamow-Teller reduction factor should lie in between 0.677 and 1.

\begin{figure}
\includegraphics[scale=1.0]{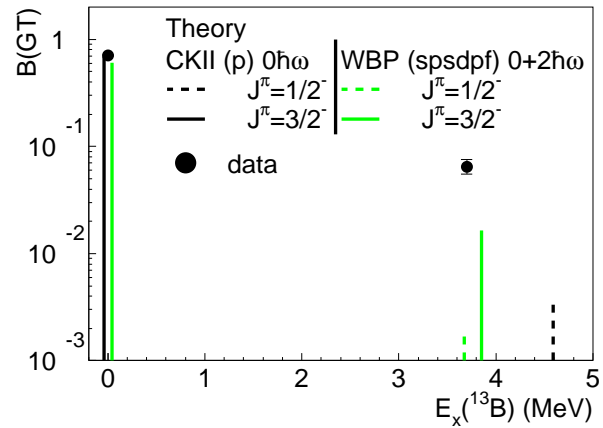}
\caption{\label{fig4}(color online) Comparison between the experimental Gamow-Teller strengths of Table \ref{tab:gt} and the theoretical values calculation with the CKII interaction ($0 \hbar\omega$) and WBP interaction (0$+$2)$\hbar\omega$. Dashed lines indicate $1/2^{-}$ states and solid lines correspond to $3/2^{-}$ states. The theoretical values have been scaled by a factor of 0.677 to account for ``quenching'' as discussed in detail in the text.}
\end{figure}

In Table \ref{tab:compare}, the theoretical predictions for the Gamow-Teller strengths are compared with those extracted from the experiment (see Table \ref{tab:gt}). This comparison is visualized in Fig. \ref{fig4}.
Besides the calculations that employ the WBP interaction with mixed (0$+$2)$\hbar\omega$ configurations as discussed above, theoretical results with the CKII interaction \cite{COH67} in $p$-shell model space ($0 \hbar\omega$ only) are also included in the table. In the latter calculation, the experimental B(GT) for the ground-state transition is well described if the reduction factor of 0.677 is applied, but the next transition is to a $1/2^{-}$ state at 4.59 MeV and its strength is more than a factor of 16 smaller than the experimentally found value.  Such a low value is below the sensitivity of the current experiment. It is, therefore, unlikely that the state at 3.6 MeV is associated with a $0 \hbar\omega$ $J^{\pi}=1/2^{-}$ state. We further note that two $3/2^{-}$ states not included in Table \ref{tab:compare} were predicted in the $0 \hbar\omega$ calculation at higher excitation energies. Their combined B(GT) (with quenching) was small, only 0.002.

In contrast to the calculation in the $0 \hbar\omega$ space, the calculation with the mixed configuration space predicts a B(GT) slightly lower (0.612) than the data (0.711(2)) for the ground-state transition after the reduction factor of 0.677 is applied. As mentioned above, this is expected because of the partial double-counting of quenching effects. The relatively strong $3/2^{-}$ state predicted at 3.86 MeV with the WBP interaction with mixed (0$+$2)$\hbar\omega$ configurations matches the location of state found experimentally at 3.6 MeV quite well, but the theoretical strength is about a factor of 3 too low, likely due to the sensitivity of the theoretical value to the exact level of mixing between $0\hbar\omega$ and $2\hbar\omega$ configurations discussed above.  The assignment of $J^{\pi}=3/2^{-}$ to the state at 3.6 MeV corresponds to the tentative assignment reported in Ref. \cite{IWA09}. The calculated decay properties of this $3/2^{-}$ state with the WBP (WBT) interaction are 7.0 (3.8) ps for the mean lifetime with B(M1)$=5.7\cdot10^{-5}$ W.u. ($1.4\cdot10^{-4}$ W.u.) and B(E2)=0.051 W.u. (0.046 W.u.), compared to the experimental values reported in Ref. \cite{IWA09} of 1.3(3) ps, B(M1)$<7.2\cdot10^{-4}$ W.u. and B(E2)$<0.81$ W.u. Thus both the B(GT) to the excited $3/2^{-}$ state obtained in the present experiment and the lifetime suggest that the mixing between $0\hbar\omega$ and $2\hbar\omega$ configurations is about a factor of 2-3 larger than obtained with the WBP and WBT interactions.

Similar to the $0 \hbar\omega$ calculation, a low-lying $1/2^{-}$ state is also predicted in the (0$+$2)$\hbar\omega$ model, albeit at a slightly lower excitation energy of 3.675 MeV. Again, the strength of this state is too low to be seen in the present experiment. Moreover, if the energy were indeed so close to that of the second $3/2^{-}$ state, the two would not be separable with the energy resolution achieved in the experiment. Finally, we found that in the mixed (0$+$2)$\hbar\omega$ calculation, many $1/2^{-}$ and $3/2^{-}$ states were predicted at excitation energies above 5 MeV as a consequence of the configuration mixing. Up to 18 MeV (which was the cut-off for the calculations), a summed B(GT) of 0.134 (0.091 after applying the reduction factor of 0.677) was distributed over nearly 100 states. None of these states was of sufficient strength to be observed and/or resolved in the current data set.

\subsection{Comparison between theoretical and experimental results for dipole transitions}
\begin{figure}
\includegraphics[scale=1.0]{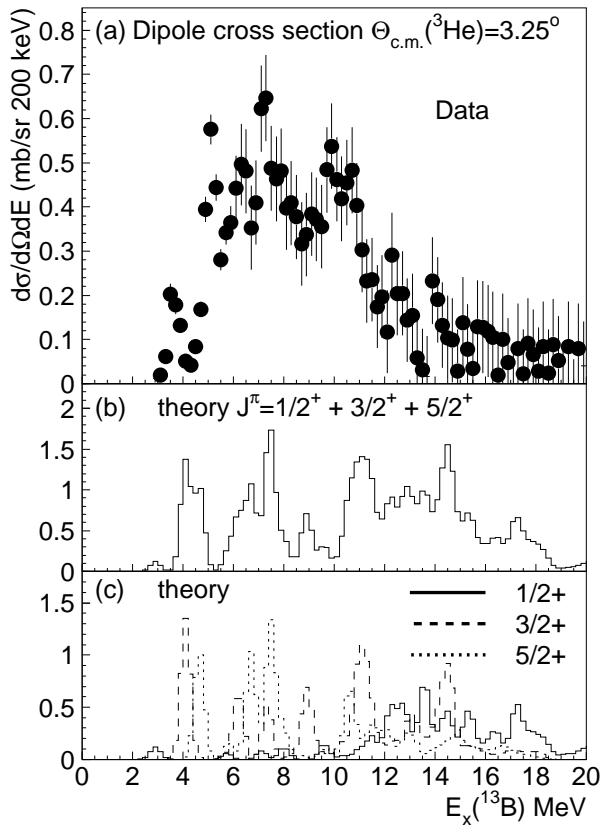}
\caption{\label{fig5}(a) Contribution of dipole transitions to the experimental excitation-energy spectrum in $^{13}$B for $3.0^{\circ}<\theta_{c.m.}<3.5^{\circ}$. This spectrum was obtained by subtracting the estimated contributions from quasifree processes for excitation energies above 5 MeV from the full spectrum (see Fig. \ref{fig3}). For the peak at 3.6 MeV, the dipole contribution was determined from the decomposition shown in Fig. \ref{fig1}(d). (b) Theoretical prediction of the dipole cross section for $3.0^{\circ}<\theta_{c.m.}<3.5^{\circ}$. The calculation is based on shell-model results which employ the adjusted WBP interaction with mixed $(1+3)\hbar\omega$ configurations and that were input for the DWBA calculation as detailed in the text. The theoretical spectrum has been folded with the experimental energy resolution.
(c) Idem, but the contributions from dipole transitions to $J^{\pi}=1/2^{+},3/2^{+},5/2^{+}$ states in $^{13}$B are separately indicated.}
\end{figure}

In this section, we focus on the comparison between the experimental and theoretical predictions of dipole transitions. For reasons discussed in Section \ref{sec:intro}, this study was performed by comparing cross sections rather than strength. In Fig. \ref{fig5}(a), the extracted collective dipole cross sections at $\theta_{c.m.}=3.25^{\circ}$ as a function of excitation energy are shown. This figure was obtained by subtracting the estimated contributions from the quasifree continuum (see Fig. \ref{fig3}) from the full spectrum at that angle. The error bars show the statistical uncertainties, combined with the estimated uncertainties in contributions of the continuum. In Fig. \ref{fig5}(b), the theoretical cross sections at $\theta_{c.m.}=3.25^{\circ}$ are shown for comparison. The curve contains contributions from transitions to $1/2^{+}$, $3/2^{+}$ and $5/2^{+}$ states, which are shown separately out in Fig. \ref{fig5}(c). These theoretical cross sections were calculated in DWBA as discussed in Section \ref{sec:results}, by using input from the shell-model calculations that employ the WBP interaction with mixed $(1+3)\hbar\omega$ configurations. For the width of the states, the experimental energy resolution was assumed (480 keV); this ignores the effects of decay on the width of the states above the 4.787 MeV threshold for decay by neutron emission. Up to about 11 MeV, the dipole cross section is mostly due to transitions to $3/2^{+}$ and $5/2^{+}$ states. Above that energy, transitions to $1/2^{+}$ states are predicted to be the strongest.

Up to about 11 MeV, the structures seen in the experimental (Fig. \ref{fig5}(a)) and theoretical (Fig. \ref{fig5}(b)) spectra are qualitatively similar. Peaks in the experimental dipole distribution are observed at 3.6 MeV (this is the dipole component of the fit shown in Fig. \ref{fig1}(c)), 5.2 MeV (Fig. \ref{fig1}(d)), 7 MeV and 10 MeV. These could possibly correspond with the maxima seen, mostly due to transitions to (combinations of) $3/2^{+}$ and $5/2^{+}$ states, in the theoretical distribution at about 4.2 MeV, 6.5 MeV, 7.5 MeV and 11 MeV, respectively. Above 11 MeV, the qualitative correspondence between data and theory is not very good: the experimental dipole distribution drops off quickly with increasing excitation energy, but significant contributions from transitions to $1/2^{+}$ states are predicted by theory.

\begin{figure}
\includegraphics[scale=1.0]{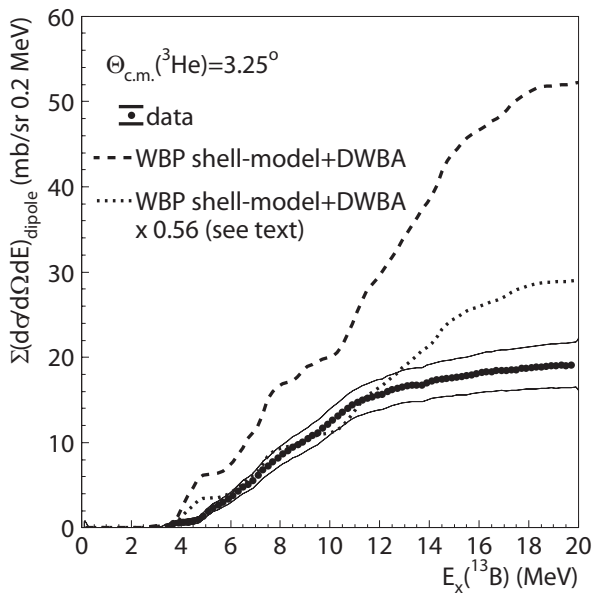}
\caption{\label{fig6} Cumulative cross sections as a function of excitation energy for dipole transitions excited via $^{13}$C($t$,$^{3}$He). The thick solid line represents the experimental result, with the thin solid line indicating the error. The dashed line represent the result from the theoretical calculations in DWBA with input from the shell-model calculation that uses the adjusted WBP interaction with mixed $(1+3)\hbar\omega$ configurations. The dotted line represents the same calculation but scaled by a factor 0.56 for reasons discussed in the text.}
\end{figure}

In Fig. \ref{fig6}, cumulative cross sections as a function of excitation energy are plotted for both the theoretical and the experimental dipole spectra at $\theta_{c.m.}=3.25^{\circ}$. The theoretical cross sections (dashed line) overshoot the data by a large factor. Two possible reasons were identified: the DWBA calculations for dipole transitions systematically overestimate the experimental cross sections and/or the dipole strengths are quenched, similar to the case for Gamow-Teller transitions.
As described in section \ref{sec:results}, the theoretical unit cross section for Gamow-Teller transitions was found to be 20\% larger than the experimental value. Under the assumption that the causes for that discrepancy affect dipole transitions as much as Gamow-Teller transitions, this percentage can be used as an estimate for the systematic error for theoretical dipole cross sections as well. We found that the effects of exchange contributions to the cross sections (estimated by comparing DWBA calculations with and without exchange) were hardly different (a few percent) for Gamow-Teller and dipole transitions, suggesting that systematic effects would be similar as well.
However, some ambiguity remained due to the use of the approximate treatment of exchange in the DWBA code. Possible differences in distortions (described by the optical model parameters) between Gamow-Teller and dipole transitions were tested by comparing cross section calculations in Plane-Wave Born Approximation (PWBA) with DWBA results. Peak Gamow-Teller and dipole cross section were reduced by a very similar factor in DWBA, compared to PWBA, suggesting that systematic errors in the optical potential would also be similar.

As an estimate for the possible effects of quenching of dipole strengths, we used the value of the reduction factor for Gamow-Teller transitions $q_{s}^{2}=(0.823)^{2}=0.677$. Combined with the 20\% decrease due to systematic errors in the DWBA calculation discussed above, we multiplied the theoretical cross sections for dipole transitions by $\frac{0.677}{1.2}=0.56$. The correspondingly scaled cumulative cross section is also shown in Fig. \ref{fig6}. A good correspondence between the magnitudes of the experimental and scaled theoretical results is found up to about 11 MeV. Above that value, the experimental curve levels off, whereas the theoretical curve continues to rise and only levels off above $\sim 18$ MeV. If the scaling factor of 0.56 for the theory is reasonably accurate, one can speculate that the discrepancy above 11 MeV is due to transitions to $1/2^{+}$ states, which, according to the theory, are strongly populated in that energy range. A possible reason could be that at these high excitation energies, couplings to configurations more complex than included in the present shell-model calculations, start playing a role and push part of the dipole strength even up to higher energy.

Although the comparison between experimental and theoretical dipole cross sections suggests that dipole strengths are reduced due to quenching by a factor similar to that found for Gamow-Teller transitions, remaining uncertainties in the DWBA and shell-model calculations make it hard to draw a definite conclusion.

\section{Conclusion}
We measured the $^{13}$C($t$,$^{3}$He) reaction at $E_{t}$=115 AMeV and extracted the Gamow-Teller strength distribution. Besides the strong transition to the $^{13}$B ground state (B(GT)=0.711(2), known from $\beta$-decay) a second Gamow-Teller transition to a state at 3.6 MeV was found, with B(GT)=0.065(5). This state very likely corresponds to the state that was recently found at 3.53 MeV in Ref. \cite{IWA09} and tentatively assigned to $J^{\pi}=3/2^{-}$. By comparing our results with shell-model calculations that employed the WBP interaction adjusted to allow for mixed $(0+2)\hbar\omega$ configurations, we confirmed that spin-parity assignment. The B(GT) predicted by the theory is somewhat lower than the data, indicating that the mixing between $0\hbar\omega$ and $2\hbar\omega$ configurations predicted in the shell model with the WBP and WBT interactions is not quite sufficient. The same conclusion is drawn by comparing the experimental results for the lifetime, B(M1) and B(E2) from Ref. \cite{IWA09} with the theory.

In order to perform shell-model calculations with the mixed $n\hbar\omega +(n+2)\hbar\omega$ ($n=0,1$) configurations, the energy of the $(n+2)\hbar\omega$ configurations was lowered to approximate the effect of $(n+4)\hbar\omega$ and higher configurations. As a consequence, the relatively large spectroscopic factor for proton transfer from $^{12}$Be to a 1/2$^{ + }$ state at 4.83 MeV in $^{13}$B \cite{OTA08} could be explained as well.

Above an excitation energy of 4 MeV, the spectrum measured via the $^{13}$C($t$,$^{3}$He) reaction is dominated by dipole transitions to states with $J^{\pi}=1/2^{+}, 3/2^{+}, 5/2^{+}$. The experimental cross sections up to 20 MeV to these states were compared with theoretical calculations. The theoretical cross sections were larger than the experimental values, but under the assumption that systematic errors in the DWBA calculation and quenching of strength are similar for dipole and Gamow-Teller transitions, a reasonable correspondence was found up to an excitation-energy of 11 MeV. Above 11 MeV, shell-model calculations predict that the dipole response is dominated by transitions to $1/2^{+}$ states, but we found little evidence for strong dipole transitions at these higher excitation energies.

\begin{acknowledgments}
We thank the cyclotron and A1900 staff at NSCL for their support during the experiment. This work was supported by the US NSF (PHY-0822648 (JINA), PHY-0606007, and PHY-0758099).
\end{acknowledgments}

\bibliography{prc}% Produces the bibliography via BibTeX.

\begin{thebibliography}{57}
\expandafter\ifx\csname natexlab\endcsname\relax\def\natexlab#1{#1}\fi
\expandafter\ifx\csname bibnamefont\endcsname\relax
  \def\bibnamefont#1{#1}\fi
\expandafter\ifx\csname bibfnamefont\endcsname\relax
  \def\bibfnamefont#1{#1}\fi
\expandafter\ifx\csname citenamefont\endcsname\relax
  \def\citenamefont#1{#1}\fi
\expandafter\ifx\csname url\endcsname\relax
  \def\url#1{\texttt{#1}}\fi
\expandafter\ifx\csname urlprefix\endcsname\relax\def\urlprefix{URL }\fi
\providecommand{\bibinfo}[2]{#2}
\providecommand{\eprint}[2][]{\url{#2}}

\bibitem[{\citenamefont{Wilkinson{\it{ et al.}}}(1959)}]{WIL59}
\bibinfo{author}{\bibfnamefont{D.}~\bibnamefont{Wilkinson{\it{ et al.}}}},
  \bibinfo{journal}{Phys. Rev.} \textbf{\bibinfo{volume}{113}},
  \bibinfo{pages}{563} (\bibinfo{year}{1959}).

\bibitem[{\citenamefont{Iwasaki{\it{ et al.}}}(2000)}]{IWA00}
\bibinfo{author}{\bibfnamefont{H.}~\bibnamefont{Iwasaki{\it{ et al.}}}},
  \bibinfo{journal}{Phys. Lett.} \textbf{\bibinfo{volume}{\bf{B481}}},
  \bibinfo{pages}{8} (\bibinfo{year}{2000}).

\bibitem[{\citenamefont{Shimoura{\it{ et al.}}}(2007)}]{SHI07}
\bibinfo{author}{\bibfnamefont{S.}~\bibnamefont{Shimoura{\it{ et al.}}}},
  \bibinfo{journal}{Phys. Lett.} \textbf{\bibinfo{volume}{\bf{B654}}},
  \bibinfo{pages}{87} (\bibinfo{year}{2007}).

\bibitem[{\citenamefont{Hamamoto and Shimoura}(2007)}]{HAM07}
\bibinfo{author}{\bibfnamefont{I.}~\bibnamefont{Hamamoto}} \bibnamefont{and}
  \bibinfo{author}{\bibfnamefont{S.}~\bibnamefont{Shimoura}},
  \bibinfo{journal}{J. Phys. G} \textbf{\bibinfo{volume}{\bf{34}}},
  \bibinfo{pages}{2714} (\bibinfo{year}{2007}).

\bibitem[{\citenamefont{Suzuki et~al.}(2003)\citenamefont{Suzuki, Fujimoto, and
  Otsuka}}]{SUZ03}
\bibinfo{author}{\bibfnamefont{T.}~\bibnamefont{Suzuki}},
  \bibinfo{author}{\bibfnamefont{R.}~\bibnamefont{Fujimoto}}, \bibnamefont{and}
  \bibinfo{author}{\bibfnamefont{T.}~\bibnamefont{Otsuka}},
  \bibinfo{journal}{Phys. Rev. C} \textbf{\bibinfo{volume}{67}},
  \bibinfo{pages}{044302} (\bibinfo{year}{2003}).

\bibitem[{\citenamefont{Ota{\it{ et al.}}}(2008)}]{OTA08}
\bibinfo{author}{\bibfnamefont{S.}~\bibnamefont{Ota{\it{ et al.}}}},
  \bibinfo{journal}{Phys. Lett.} \textbf{\bibinfo{volume}{\bf{B666}}},
  \bibinfo{pages}{311} (\bibinfo{year}{2008}).

\bibitem[{\citenamefont{Iwasaki et~al.}(2009)}]{IWA09}
\bibinfo{author}{\bibfnamefont{H.}~\bibnamefont{Iwasaki}} \bibnamefont{et~al.},
  \bibinfo{journal}{Phys. Rev. Lett.} \textbf{\bibinfo{volume}{\bf{102}}},
  \bibinfo{pages}{202502} (\bibinfo{year}{2009}).

\bibitem[{\citenamefont{Harakeh and van~der Woude}(2001)}]{HAR01}
\bibinfo{author}{\bibfnamefont{M.~N.} \bibnamefont{Harakeh}} \bibnamefont{and}
  \bibinfo{author}{\bibfnamefont{A.}~\bibnamefont{van~der Woude}},
  \emph{\bibinfo{title}{Giant Resonances: Fundamental High-Frequency Modes of
  Nuclear Excitations}} (\bibinfo{publisher}{Oxford University Press},
  \bibinfo{address}{New York}, \bibinfo{year}{2001}).

\bibitem[{\citenamefont{Osterfeld}(1992)}]{OST92}
\bibinfo{author}{\bibfnamefont{F.}~\bibnamefont{Osterfeld}},
  \bibinfo{journal}{Rev. Mod. Phys.} \textbf{\bibinfo{volume}{64}},
  \bibinfo{pages}{491} (\bibinfo{year}{1992}).

\bibitem[{\citenamefont{Taddeucci{\it{ et al.}}}(1987)}]{TAD87}
\bibinfo{author}{\bibfnamefont{T.~D.} \bibnamefont{Taddeucci{\it{ et al.}}}},
  \bibinfo{journal}{Nucl. Phys.} \textbf{\bibinfo{volume}{\bf{A469}}},
  \bibinfo{pages}{125} (\bibinfo{year}{1987}).

\bibitem[{\citenamefont{Ajzenberg-Selove}(1991)}]{AJZ91}
\bibinfo{author}{\bibfnamefont{F.}~\bibnamefont{Ajzenberg-Selove}},
  \bibinfo{journal}{Nucl. Phys.} \textbf{\bibinfo{volume}{\bf{A523}}},
  \bibinfo{pages}{1} (\bibinfo{year}{1991}).

\bibitem[{\citenamefont{Zegers{\it{ et al.}}}(2006)}]{ZEG06}
\bibinfo{author}{\bibfnamefont{R.~G.~T.} \bibnamefont{Zegers{\it{ et al.}}}},
  \bibinfo{journal}{Phys. Rev. C} \textbf{\bibinfo{volume}{\bf{74}}},
  \bibinfo{pages}{024309} (\bibinfo{year}{2006}).

\bibitem[{\citenamefont{Cole{\it{ et al.}}}(2006)}]{COL06}
\bibinfo{author}{\bibfnamefont{A.~L.} \bibnamefont{Cole{\it{ et al.}}}},
  \bibinfo{journal}{Phys. Rev. C} \textbf{\bibinfo{volume}{\bf{74}}},
  \bibinfo{pages}{034333} (\bibinfo{year}{2006}).

\bibitem[{\citenamefont{Fujita{\it{ et al.}}}(2007)}]{FUJ07}
\bibinfo{author}{\bibfnamefont{Y.}~\bibnamefont{Fujita{\it{ et al.}}}},
  \bibinfo{journal}{Phys. Rev. C} \textbf{\bibinfo{volume}{\bf{75}}},
  \bibinfo{pages}{057305} (\bibinfo{year}{2007}).

\bibitem[{\citenamefont{Zegers{\it{ et al.}}}(2007)}]{ZEG07}
\bibinfo{author}{\bibfnamefont{R.~G.~T.} \bibnamefont{Zegers{\it{ et al.}}}},
  \bibinfo{journal}{Phys. Rev. Lett.} \textbf{\bibinfo{volume}{\bf{99}}},
  \bibinfo{pages}{202501} (\bibinfo{year}{2007}).

\bibitem[{\citenamefont{Zegers{\it{ et al.}}}(2008)}]{ZEG08}
\bibinfo{author}{\bibfnamefont{R.~G.~T.} \bibnamefont{Zegers{\it{ et al.}}}},
  \bibinfo{journal}{Phys. Rev. C} \textbf{\bibinfo{volume}{\bf{77}}},
  \bibinfo{pages}{024307} (\bibinfo{year}{2008}).

\bibitem[{\citenamefont{Dmitriev et~al.}(2001)\citenamefont{Dmitriev,
  Zelevinsky, and Austin}}]{DMI01}
\bibinfo{author}{\bibfnamefont{V.~F.} \bibnamefont{Dmitriev}},
  \bibinfo{author}{\bibfnamefont{V.}~\bibnamefont{Zelevinsky}},
  \bibnamefont{and} \bibinfo{author}{\bibfnamefont{S.~M.}
  \bibnamefont{Austin}}, \bibinfo{journal}{Phys. Rev. C}
  \textbf{\bibinfo{volume}{\bf{65}}}, \bibinfo{pages}{015803}
  (\bibinfo{year}{2001}).

\bibitem[{\citenamefont{Yako et~al.}(2006)\citenamefont{Yako, Sagawa, and
  Sakai}}]{YAK06}
\bibinfo{author}{\bibfnamefont{K.}~\bibnamefont{Yako}},
  \bibinfo{author}{\bibfnamefont{H.}~\bibnamefont{Sagawa}}, \bibnamefont{and}
  \bibinfo{author}{\bibfnamefont{H.}~\bibnamefont{Sakai}},
  \bibinfo{journal}{Phys. Rev. C} \textbf{\bibinfo{volume}{\bf{74}}},
  \bibinfo{pages}{051303(R)} (\bibinfo{year}{2006}).

\bibitem[{CCF()}]{CCF}
\bibinfo{note}{The K500$\otimes$K1200, a coupled cyclotron facility at the
  NSCL, NSCL Report MSUCL-939}.

\bibitem[{\citenamefont{Hitt{\it{ et al.}}}(2006)}]{HIT06}
\bibinfo{author}{\bibfnamefont{G.~W.} \bibnamefont{Hitt{\it{ et al.}}}},
  \bibinfo{journal}{Nucl. Instrum. Methods Phys. Res. A}
  \textbf{\bibinfo{volume}{\bf{566}}}, \bibinfo{pages}{264}
  (\bibinfo{year}{2006}).

\bibitem[{\citenamefont{Morrissey{\it{ et al.}}}(2003)}]{MOR03}
\bibinfo{author}{\bibfnamefont{D.}~\bibnamefont{Morrissey{\it{ et al.}}}},
  \bibinfo{journal}{Nucl. Instrum. Meth. Phys. Res. B}
  \textbf{\bibinfo{volume}{\bf{204}}}, \bibinfo{pages}{90}
  (\bibinfo{year}{2003}).

\bibitem[{\citenamefont{Bazin et~al.}(2003)\citenamefont{Bazin, Caggiano,
  Sherrill, Yurkon, and Zeller}}]{BAZ03}
\bibinfo{author}{\bibfnamefont{D.}~\bibnamefont{Bazin}},
  \bibinfo{author}{\bibfnamefont{J.~A.} \bibnamefont{Caggiano}},
  \bibinfo{author}{\bibfnamefont{B.~M.} \bibnamefont{Sherrill}},
  \bibinfo{author}{\bibfnamefont{J.}~\bibnamefont{Yurkon}}, \bibnamefont{and}
  \bibinfo{author}{\bibfnamefont{A.}~\bibnamefont{Zeller}},
  \bibinfo{journal}{Nucl. Instr. Meth. Phys. Res. B}
  \textbf{\bibinfo{volume}{\bf{204}}}, \bibinfo{pages}{629}
  (\bibinfo{year}{2003}).

\bibitem[{\citenamefont{Howard{\it{ et al.}}}(2008)}]{HOW08}
\bibinfo{author}{\bibfnamefont{M.}~\bibnamefont{Howard{\it{ et al.}}}},
  \bibinfo{journal}{Phys. Rev. C} \textbf{\bibinfo{volume}{\bf{78}}},
  \bibinfo{pages}{047302} (\bibinfo{year}{2008}).

\bibitem[{\citenamefont{Hitt{\it{ et al.}}}(2009)}]{HIT09}
\bibinfo{author}{\bibfnamefont{G.}~\bibnamefont{Hitt{\it{ et al.}}}},
  \bibinfo{journal}{Phys. Rev. C}  (\bibinfo{year}{2009}), \bibinfo{note}{in
  press}.

\bibitem[{\citenamefont{Noji}(2008)}]{NOJ08}
\bibinfo{author}{\bibfnamefont{S.}~\bibnamefont{Noji}}, in
  \emph{\bibinfo{booktitle}{Future Prospects for Spectroscopy and Direct
  Reactions}} (\bibinfo{year}{2008}), pp. \bibinfo{pages}{8,9},
  \bibinfo{note}{http://meetings.nscl.msu.edu/fp2008/presentations/noji.pdf}.

\bibitem[{\citenamefont{Yurkon et~al.}(1999)\citenamefont{Yurkon, Bazin,
  Benenson, Morrissey, Sherrill, Swan, and Swanson}}]{YUR99}
\bibinfo{author}{\bibfnamefont{J.}~\bibnamefont{Yurkon}},
  \bibinfo{author}{\bibfnamefont{D.}~\bibnamefont{Bazin}},
  \bibinfo{author}{\bibfnamefont{W.}~\bibnamefont{Benenson}},
  \bibinfo{author}{\bibfnamefont{D.~J.} \bibnamefont{Morrissey}},
  \bibinfo{author}{\bibfnamefont{B.~M.} \bibnamefont{Sherrill}},
  \bibinfo{author}{\bibfnamefont{D.}~\bibnamefont{Swan}}, \bibnamefont{and}
  \bibinfo{author}{\bibfnamefont{R.}~\bibnamefont{Swanson}},
  \bibinfo{journal}{Nucl. Instr. Meth. Phys. Res. A}
  \textbf{\bibinfo{volume}{\bf{422}}}, \bibinfo{pages}{291}
  (\bibinfo{year}{1999}).

\bibitem[{\citenamefont{Berz et~al.}(1993)\citenamefont{Berz, Joh, Nolen,
  Sherrill, and Zeller}}]{BER93}
\bibinfo{author}{\bibfnamefont{M.}~\bibnamefont{Berz}},
  \bibinfo{author}{\bibfnamefont{K.}~\bibnamefont{Joh}},
  \bibinfo{author}{\bibfnamefont{J.~A.} \bibnamefont{Nolen}},
  \bibinfo{author}{\bibfnamefont{B.~M.} \bibnamefont{Sherrill}},
  \bibnamefont{and} \bibinfo{author}{\bibfnamefont{A.~F.}
  \bibnamefont{Zeller}}, \bibinfo{journal}{Phys. Rev. C}
  \textbf{\bibinfo{volume}{\bf{47}}}, \bibinfo{pages}{537}
  (\bibinfo{year}{1993}).

\bibitem[{\citenamefont{Matsuoka et~al.}(1980)}]{MAT80}
\bibinfo{author}{\bibfnamefont{N.}~\bibnamefont{Matsuoka}}
  \bibnamefont{et~al.}, \bibinfo{journal}{Nucl. Phys.}
  \textbf{\bibinfo{volume}{\bf{A337}}}, \bibinfo{pages}{269}
  (\bibinfo{year}{1980}).

\bibitem[{\citenamefont{Aarts et~al.}(1981)}]{AAR81}
\bibinfo{author}{\bibfnamefont{E.~H.~L.} \bibnamefont{Aarts}}
  \bibnamefont{et~al.}, \bibinfo{journal}{Phys. Lett.}
  \textbf{\bibinfo{volume}{\bf{B102}}}, \bibinfo{pages}{307}
  (\bibinfo{year}{1981}).

\bibitem[{\citenamefont{Cook and Carr}(1988)}]{FOLD}
\bibinfo{author}{\bibfnamefont{J.}~\bibnamefont{Cook}} \bibnamefont{and}
  \bibinfo{author}{\bibfnamefont{J.}~\bibnamefont{Carr}}
  (\bibinfo{year}{1988}), \bibinfo{note}{computer program \textsc{fold},
  Florida State University (unpublished), based on F. Petrovich and D. Stanley,
  Nucl. Phys. {\bf{A275}}, 487 (1977), modified as described in J. Cook {\it{
  et al.}}, Phys. Rev. C {\bf{30}}, 1538 (1984) and R. G. T. Zegers, S.
  Fracasso and G. Col\`{o} (2006), unpublished.}

\bibitem[{\citenamefont{Love and Franey}(1981)}]{LOV81}
\bibinfo{author}{\bibfnamefont{W.~G.} \bibnamefont{Love}} \bibnamefont{and}
  \bibinfo{author}{\bibfnamefont{M.~A.} \bibnamefont{Franey}},
  \bibinfo{journal}{Phys. Rev. C} \textbf{\bibinfo{volume}{\bf{24}}},
  \bibinfo{pages}{1073} (\bibinfo{year}{1981}).

\bibitem[{\citenamefont{Franey and Love}(1985)}]{LOV85}
\bibinfo{author}{\bibfnamefont{M.~A.} \bibnamefont{Franey}} \bibnamefont{and}
  \bibinfo{author}{\bibfnamefont{W.~G.} \bibnamefont{Love}},
  \bibinfo{journal}{Phys. Rev. C} \textbf{\bibinfo{volume}{\bf{31}}},
  \bibinfo{pages}{488} (\bibinfo{year}{1985}).

\bibitem[{\citenamefont{Pieper and Wiringa}(2001)}]{WIR05}
\bibinfo{author}{\bibfnamefont{S.~C.} \bibnamefont{Pieper}} \bibnamefont{and}
  \bibinfo{author}{\bibfnamefont{R.~B.} \bibnamefont{Wiringa}},
  \bibinfo{journal}{Annu. Rev. Nucl. Part. Sci.}
  \textbf{\bibinfo{volume}{\bf{51}}}, \bibinfo{pages}{53}
  (\bibinfo{year}{2001}), \bibinfo{note}{and R.B. Wiringa, private
  communication}.

\bibitem[{\citenamefont{Brown{\it{ et al.}}}()}]{OXBA}
\bibinfo{author}{\bibfnamefont{B.~A.} \bibnamefont{Brown{\it{ et al.}}}},
  \bibinfo{note}{{N}SCL report MSUCL-1289}.

\bibitem[{\citenamefont{Warburton and Brown}(1992)}]{WAR92}
\bibinfo{author}{\bibfnamefont{E.~K.} \bibnamefont{Warburton}}
  \bibnamefont{and} \bibinfo{author}{\bibfnamefont{B.~A.} \bibnamefont{Brown}},
  \bibinfo{journal}{Phys. Rev. C} \textbf{\bibinfo{volume}{\bf{46}}},
  \bibinfo{pages}{923} (\bibinfo{year}{1992}).

\bibitem[{\citenamefont{Brown}(1998)}]{BRO98}
\bibinfo{author}{\bibfnamefont{B.~A.} \bibnamefont{Brown}},
  \bibinfo{journal}{Phys. Rev. C} \textbf{\bibinfo{volume}{\bf{58}}},
  \bibinfo{pages}{220} (\bibinfo{year}{1998}).

\bibitem[{\citenamefont{Fujimura{\it{ et al.}}}(2004)}]{FUJ04}
\bibinfo{author}{\bibfnamefont{H.}~\bibnamefont{Fujimura{\it{ et al.}}}},
  \bibinfo{journal}{Phys. Rev. C} \textbf{\bibinfo{volume}{\bf{69}}},
  \bibinfo{pages}{064327} (\bibinfo{year}{2004}), \bibinfo{note}{and private
  communication.}

\bibitem[{\citenamefont{van~der Werf et~al.}(1989)\citenamefont{van~der Werf,
  Brandenburg, Grasdijk, Sterrenburg, Harakeh, Greenfield, Brown, and
  Fujiwara}}]{WER89}
\bibinfo{author}{\bibfnamefont{S.~Y.} \bibnamefont{van~der Werf}},
  \bibinfo{author}{\bibfnamefont{S.}~\bibnamefont{Brandenburg}},
  \bibinfo{author}{\bibfnamefont{P.}~\bibnamefont{Grasdijk}},
  \bibinfo{author}{\bibfnamefont{W.~A.} \bibnamefont{Sterrenburg}},
  \bibinfo{author}{\bibfnamefont{M.~N.} \bibnamefont{Harakeh}},
  \bibinfo{author}{\bibfnamefont{M.~B.} \bibnamefont{Greenfield}},
  \bibinfo{author}{\bibfnamefont{B.~A.} \bibnamefont{Brown}}, \bibnamefont{and}
  \bibinfo{author}{\bibfnamefont{M.}~\bibnamefont{Fujiwara}},
  \bibinfo{journal}{Nucl. Phys.} \textbf{\bibinfo{volume}{\bf{A496}}},
  \bibinfo{pages}{305} (\bibinfo{year}{1989}).

\bibitem[{\citenamefont{Udagawa et~al.}(1987)\citenamefont{Udagawa, Schulte,
  and Osterfeld}}]{UDA87}
\bibinfo{author}{\bibfnamefont{T.}~\bibnamefont{Udagawa}},
  \bibinfo{author}{\bibfnamefont{A.}~\bibnamefont{Schulte}}, \bibnamefont{and}
  \bibinfo{author}{\bibfnamefont{F.}~\bibnamefont{Osterfeld}},
  \bibinfo{journal}{Nucl. Phys.} \textbf{\bibinfo{volume}{\bf{A474}}},
  \bibinfo{pages}{131} (\bibinfo{year}{1987}).

\bibitem[{\citenamefont{Kim et~al.}(2000)\citenamefont{Kim, Knobles, Stotts,
  and Udagawa}}]{KIM00}
\bibinfo{author}{\bibfnamefont{B.~T.} \bibnamefont{Kim}},
  \bibinfo{author}{\bibfnamefont{D.~P.} \bibnamefont{Knobles}},
  \bibinfo{author}{\bibfnamefont{S.~A.} \bibnamefont{Stotts}},
  \bibnamefont{and} \bibinfo{author}{\bibfnamefont{T.}~\bibnamefont{Udagawa}},
  \bibinfo{journal}{Phys. Rev. C} \textbf{\bibinfo{volume}{\bf{61}}},
  \bibinfo{pages}{044611} (\bibinfo{year}{2000}).

\bibitem[{\citenamefont{Nakayama et~al.}(1990)}]{NAK90}
\bibinfo{author}{\bibfnamefont{S.}~\bibnamefont{Nakayama}}
  \bibnamefont{et~al.}, \bibinfo{journal}{Nucl. Phys.}
  \textbf{\bibinfo{volume}{\bf{A507}}}, \bibinfo{pages}{515}
  (\bibinfo{year}{1990}).

\bibitem[{\citenamefont{Guimaraes et~al.}(2000)}]{GUI00}
\bibinfo{author}{\bibfnamefont{V.}~\bibnamefont{Guimaraes}}
  \bibnamefont{et~al.}, \bibinfo{journal}{Phys. Rev. C}
  \textbf{\bibinfo{volume}{\bf{61}}}, \bibinfo{pages}{064609}
  (\bibinfo{year}{2000}).

\bibitem[{\citenamefont{Middleton and Pullen}(1964)}]{MID64}
\bibinfo{author}{\bibfnamefont{R.}~\bibnamefont{Middleton}} \bibnamefont{and}
  \bibinfo{author}{\bibfnamefont{D.~J.} \bibnamefont{Pullen}},
  \bibinfo{journal}{Nucl. Phys.} \textbf{\bibinfo{volume}{\bf{51}}},
  \bibinfo{pages}{50} (\bibinfo{year}{1964}).

\bibitem[{\citenamefont{Wyborny}(1972)}]{WYB72}
\bibinfo{author}{\bibfnamefont{H.~W.} \bibnamefont{Wyborny}},
  \bibinfo{journal}{Nucl. Phys.} \textbf{\bibinfo{volume}{\bf{A185}}},
  \bibinfo{pages}{669} (\bibinfo{year}{1972}).

\bibitem[{\citenamefont{Ajzenberg-Selove
  et~al.}(1978)\citenamefont{Ajzenberg-Selove, Flynn, and Hansen}}]{AZJ78}
\bibinfo{author}{\bibfnamefont{F.}~\bibnamefont{Ajzenberg-Selove}},
  \bibinfo{author}{\bibfnamefont{E.~R.} \bibnamefont{Flynn}}, \bibnamefont{and}
  \bibinfo{author}{\bibfnamefont{O.}~\bibnamefont{Hansen}},
  \bibinfo{journal}{Phys. Rev. C} \textbf{\bibinfo{volume}{\bf{17}}},
  \bibinfo{pages}{1283} (\bibinfo{year}{1978}).

\bibitem[{\citenamefont{J\"{a}necke et~al.}(1993)}]{JAN93}
\bibinfo{author}{\bibfnamefont{J.}~\bibnamefont{J\"{a}necke}}
  \bibnamefont{et~al.}, \bibinfo{journal}{Phys. Rev. C}
  \textbf{\bibinfo{volume}{48}}, \bibinfo{pages}{2828} (\bibinfo{year}{1993}).

\bibitem[{\citenamefont{Erell et~al.}(1986)}]{ERE86}
\bibinfo{author}{\bibfnamefont{A.}~\bibnamefont{Erell}} \bibnamefont{et~al.},
  \bibinfo{journal}{Phys. Rev. C} \textbf{\bibinfo{volume}{34}},
  \bibinfo{pages}{1822} (\bibinfo{year}{1986}).

\bibitem[{\citenamefont{Borghols}(1988)}]{BOR88}
\bibinfo{author}{\bibfnamefont{W.~T.~A.} \bibnamefont{Borghols}},
  \bibinfo{type}{Ph.d. thesis}, \bibinfo{school}{Rijksuniversiteit Groningen}
  (\bibinfo{year}{1988}).

\bibitem[{\citenamefont{Zegers}(1999)}]{ZEG99}
\bibinfo{author}{\bibfnamefont{R.~G.~T.} \bibnamefont{Zegers}},
  \bibinfo{type}{Ph.d. thesis}, \bibinfo{school}{Rijksuniversiteit Groningen}
  (\bibinfo{year}{1999}).

\bibitem[{\citenamefont{Navin et~al.}(2000)}]{NAV00}
\bibinfo{author}{\bibfnamefont{A.}~\bibnamefont{Navin}} \bibnamefont{et~al.},
  \bibinfo{journal}{Phys. Rev. Lett.} \textbf{\bibinfo{volume}{\bf{85}}},
  \bibinfo{pages}{266} (\bibinfo{year}{2000}).

\bibitem[{\citenamefont{Warburton et~al.}(1992)\citenamefont{Warburton, Brown,
  and Millener}}]{WAR92a}
\bibinfo{author}{\bibfnamefont{E.~K.} \bibnamefont{Warburton}},
  \bibinfo{author}{\bibfnamefont{B.~A.} \bibnamefont{Brown}}, \bibnamefont{and}
  \bibinfo{author}{\bibfnamefont{D.~J.} \bibnamefont{Millener}},
  \bibinfo{journal}{Phys. Lett.} \textbf{\bibinfo{volume}{\bf{B293}}},
  \bibinfo{pages}{7} (\bibinfo{year}{1992}).

\bibitem[{\citenamefont{Snover et~al.}(1983)\citenamefont{Snover, Adelberger,
  Ikossi, and Brown}}]{SNO83}
\bibinfo{author}{\bibfnamefont{K.~A.} \bibnamefont{Snover}},
  \bibinfo{author}{\bibfnamefont{E.~G.} \bibnamefont{Adelberger}},
  \bibinfo{author}{\bibfnamefont{P.~G.} \bibnamefont{Ikossi}},
  \bibnamefont{and} \bibinfo{author}{\bibfnamefont{B.~A.} \bibnamefont{Brown}},
  \bibinfo{journal}{Phys. Rev. C} \textbf{\bibinfo{volume}{\bf{27}}},
  \bibinfo{pages}{1837} (\bibinfo{year}{1983}).

\bibitem[{\citenamefont{Chou et~al.}(1993)\citenamefont{Chou, Warburton, and
  Brown}}]{CHO93}
\bibinfo{author}{\bibfnamefont{W.-T.} \bibnamefont{Chou}},
  \bibinfo{author}{\bibfnamefont{E.~K.} \bibnamefont{Warburton}},
  \bibnamefont{and} \bibinfo{author}{\bibfnamefont{B.~A.} \bibnamefont{Brown}},
  \bibinfo{journal}{Phys. Rev. C} \textbf{\bibinfo{volume}{\bf{47}}},
  \bibinfo{pages}{163} (\bibinfo{year}{1993}).

\bibitem[{\citenamefont{Hyuga et~al.}(1980)\citenamefont{Hyuga, Arima, and
  Shimizu}}]{HYU80}
\bibinfo{author}{\bibfnamefont{H.}~\bibnamefont{Hyuga}},
  \bibinfo{author}{\bibfnamefont{A.}~\bibnamefont{Arima}}, \bibnamefont{and}
  \bibinfo{author}{\bibfnamefont{K.}~\bibnamefont{Shimizu}},
  \bibinfo{journal}{Nucl. Phys.} \textbf{\bibinfo{volume}{\bf{A336}}},
  \bibinfo{pages}{363} (\bibinfo{year}{1980}).

\bibitem[{\citenamefont{Arima}(1999)}]{ARI99}
\bibinfo{author}{\bibfnamefont{A.}~\bibnamefont{Arima}},
  \bibinfo{journal}{Nucl. Phys.} \textbf{\bibinfo{volume}{\bf{A649}}},
  \bibinfo{pages}{260c} (\bibinfo{year}{1999}).

\bibitem[{\citenamefont{Brown and Wildenthal}(1988)}]{BRO88}
\bibinfo{author}{\bibfnamefont{B.~A.} \bibnamefont{Brown}} \bibnamefont{and}
  \bibinfo{author}{\bibfnamefont{B.~H.} \bibnamefont{Wildenthal}},
  \bibinfo{journal}{Annu. Rev. Nucl. Part. Sci.}
  \textbf{\bibinfo{volume}{\bf{38}}}, \bibinfo{pages}{29}
  (\bibinfo{year}{1988}).

\bibitem[{\citenamefont{Cohen and Kurath}(1967)}]{COH67}
\bibinfo{author}{\bibfnamefont{S.}~\bibnamefont{Cohen}} \bibnamefont{and}
  \bibinfo{author}{\bibfnamefont{D.}~\bibnamefont{Kurath}},
  \bibinfo{journal}{Nucl. Phys.} \textbf{\bibinfo{volume}{\bf{A101}}},
  \bibinfo{pages}{1} (\bibinfo{year}{1967}).

\end{thebibliography}

\end{document}